\documentstyle{l-aa}
\begin{document}

\thesaurus{04.03.1; 11.19.2; 13.19.1}

\title{Global H\,{\sc i}  profiles of spiral galaxies}

\author{J.J. Kamphuis \inst{}, D. Sijbring \inst{} and T.S. van Albada
        \inst{}}

\offprints{J.J. Kamphuis}

\institute{Kapteyn Astronomical Institute, P.O. Box 800, NL-9700 AV
Groningen, The Netherlands}

\date{received March 27, 1995; accepted September 4, 1995}

\maketitle

\begin{abstract}

In this paper we present short H\,{\sc i} synthesis observations of 57
galaxies without H\,{\sc i} information in the RC3.  These are a
by-product of a large survey with the WSRT of the neutral hydrogen gas
in spiral and irregular galaxies.  Global profiles and related
quantities are given for the 42 detected galaxies and upper limits for
the remaining 15. A number of galaxies have  low values of
H\,{\sc i} mass-to-blue luminosity ratio.


\end{abstract}

\section{Introduction}

At the Kapteyn Institute a long-term project has recently been initiated
to map neutral hydrogen gas (H\,{\sc i}) in many hundreds of galaxies
with the Westerbork Synthesis Radio Telescope (WSRT).  This project,
denoted as WHISP (Westerbork observations of neutral Hydrogen in
Irregular and SPiral galaxies), will concentrate on nearby galaxies
north of declination $\delta$(1950) = 20$^{\rm o}$.
  Here we present the  results of a side-product
of WHISP: short observations of RC3 galaxies for
which no H\,{\sc i} data appear to be available.
The main aim of these short observations
is to determine the H\,{\sc i} content.
Observations of a few hours integration time are sufficient for that purpose.
 Below we present
global profiles, and parameters derived from these, for 42 galaxies.

\section{The sample}

The sample from which WHISP candidates are being selected
consists of galaxies in
the Uppsala General Catalogue of Galaxies (UGC, Nilson 1973) with blue
major diameters $d_{\rm b}$ $>$ 1.5 arcmin and
$\delta$(1950) $>$ 20$^{\rm o}$, 3148 in total.  A subsample of 200 galaxies
with
$d_{\rm b}$ $>$ 2.0 arcmin and flux densities at 21-cm larger than 200
mJy (as calculated from the ratio of total
H\,{\sc i} fluxes and profile widths listed in
the Third Reference Catalogue of Bright Galaxies (RC3, de Vaucouleurs et
al.  1991)) presently serves as the main observing list for WHISP.
Galaxies satisfying these selection criteria generally have redshifts
less than 2000 km s$^{-1}$.

As a side-product of WHISP we present short observations of
UGC galaxies, which  obey the following criteria: $d_{\rm b}$ $>$ 1.5 arcmin,
$\delta$(1950) $>$ 20$^{\rm o}$, morphological type: S0 or later, $V_{\rm
opt}$ $<$ 5000 km s$^{-1}$, and without H\,{\sc i} information in the RC3.
The latter
sample consists of 280 galaxies of which 57 have been observed.

\section{Observations and data reduction}

The observations have been carried out in the periods July - September
1992, August - September 1993 and December 1993 - February 1994 with the
WSRT, using the Digital eXtended Backend (DXB) with 64 or 128 channels
and with a total bandwidth of 1.25, 2.5 or 5 MHz; a uniform velocity
taper has been used.  The duration of the observations varies from two
to six hours.  The centre of the observed field lies within one arcmin
of the optical position of the galaxy and the centre of the spectral
passband is based on the redshift of the galaxy given in the RC3.  A
standard calibration is applied by observing 3C sources before and after
the measurement of the galaxies.

The {\it uv}-data have been Fourier transformed, with a Gaussian taper
of 390 meter (half-width), using the Dwingeloo reduction package NEWSTAR.
A weighting proportional to the track length
covered in the {\it uv}-plane has been applied and the {\it uv}-data have
been convolved with an exponential
sinc onto a rectangular grid.  This results in channel maps which contain
essentially one-dimensional information with
an  angular resolution of 60$^{\prime\prime}$ in the direction of the
resolution axis, i.e. the projection of the E-W baseline onto the sky.

\begin{figure*}
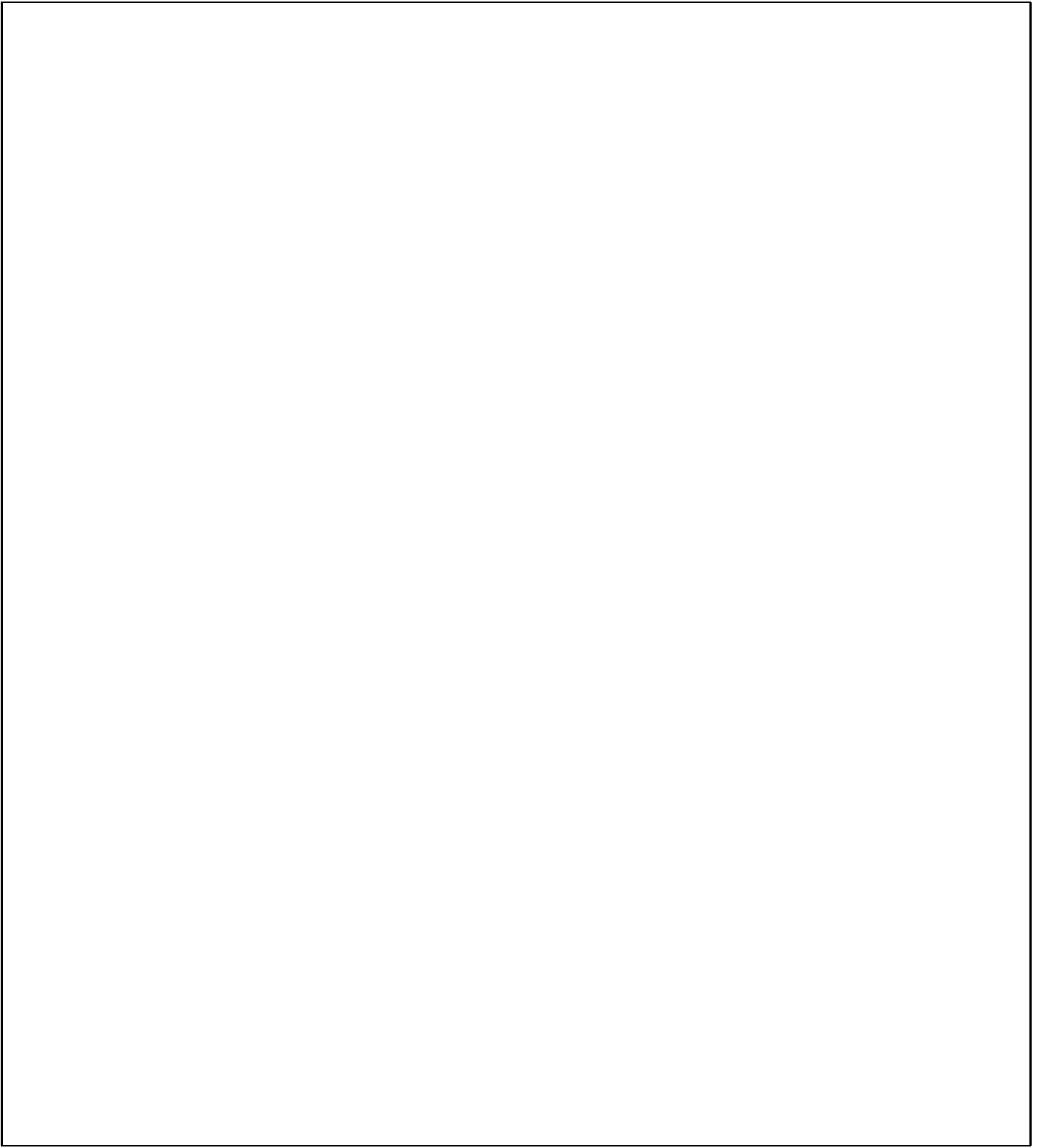

\picplace{20 cm}
\caption[]{Global H\,{\sc i} profiles for  42 galaxies.}
\end{figure*}

\addtocounter{figure}{-1}
\begin{figure*}
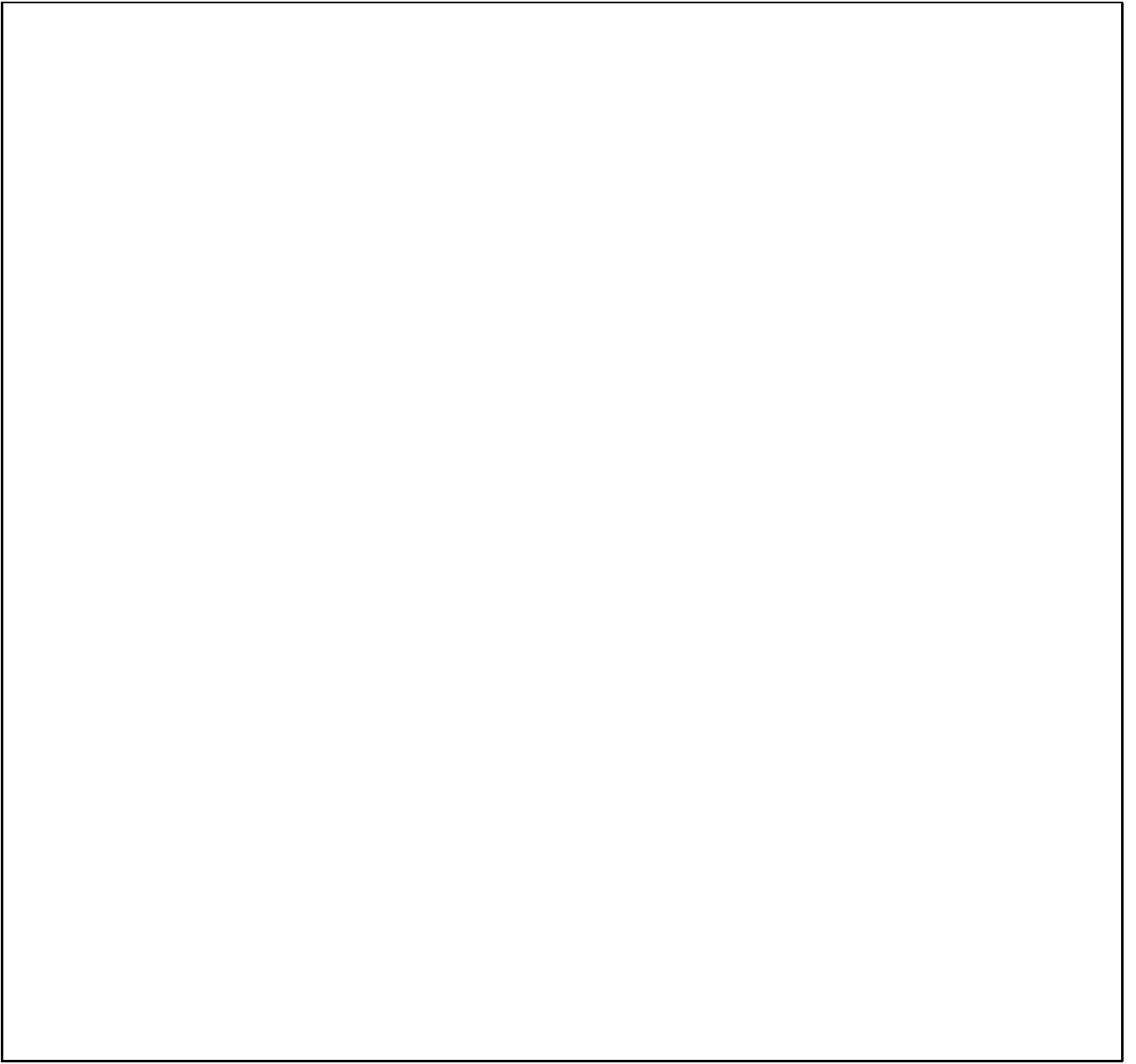

\picplace{17 cm}
\caption[]{Continued.}
\end{figure*}

The reduction of the Fourier-transformed data has been carried out with the
Groningen Image Processing SYstem (GIPSY, van der Hulst et al. 1992).
We follow the method used by Warmels (1988).
First, we have applied a Hanning smoothing in velocity to improve the
sensitivity.  Next, trial position-velocity (PV) maps along the
resolution axis have been made using a range of sizes for the
interval over which the signal is integrated in the
direction
perpendicular to the resolution axis.  The continuum emission has been
removed in the PV-maps by fitting a linear baseline in velocity to the signal
in the
channels free from H\,{\sc i} emission, subtracting the result at all
positions in the sky.  The PV-map with the highest signal-to-noise ratio
is then selected as the most appropriate map to derive the global H\,{\sc i}
properties of the galaxy.  (Note that the PV-map has only been searched
for H\,{\sc i} emission near the position and redshift of the galaxy.) Due
to the incomplete {\it uv}-coverage, the antenna pattern suffers from
sidelobes with values up to 30 - 40$\%$ of the central peak, giving
large depressions in the PV-map.  We have suppressed these effects by
CLEANing the map (H\"{o}gbom 1974) with the corresponding antenna
pattern, which is determined in the same way as the PV-map.  The r.m.s.
noise in the `cleaned' PV-maps is typically 3 to 8 mJy/beam; a  slightly
varying zero-flux baseline is visible at a low level in some noisy
profiles.

\section{Global profiles}

Because the spatial
information on the galaxies is  one-dimensional in a rather
arbitrary direction we only derive the global profiles,
i.e.  the flux densities as a function of heliocentric radial velocity.

The shortest baseline is 36, 54 or 72 meter, implying that the
observations are less sensitive to structures with sizes larger than 10,
7.5 and 5 arcmin, respectively.  Assuming that the H\,{\sc i} diameters
are, on average, about a factor of two larger than the optical sizes,
the H\,{\sc i} sizes for the galaxies in our sample will typically be
less than about 6 arc min.  Therefore, the amount of H\,{\sc i} emission
missed in the global profiles will be small.  Results for a comparison
sample, consisting of galaxies with total H\,{\sc i} fluxes known from single
dish observations of which short WSRT observations have also been taken,
indeed show that the total WSRT H\,{\sc i} fluxes are within 10 percent
of those derived from the single dish observations.  This agrees with the
findings of Warmels (1988),   Oosterloo
\& Shostak (1993) and Broeils \& van Woerden (1994).

The global profiles are shown in Figure 1; 42 out of 57 galaxies
observed have been detected.  The flux density at each radial velocity
is determined as the sum of the flux in the CLEAN components and the
flux in the residuals.  Outside the velocity range of the H\,{\sc i}
emission the flux densities are set to zero.  Several galaxies have an
H\,{\sc i} flux just above the detection limit, resulting in irregular,
noisy profiles.  A number of galaxies have profiles with peak flux
densities in excess of 100 mJy.

{}From the global profiles we have derived the total H\,{\sc i} mass, the
profile
width and the systemic velocity.  The profile width is determined as the
difference between the velocities at the low and high velocity sides of the
profile, at the customary flux density levels of 20 and 50 percent of
the peak value.  Following the precepts of Sullivan et al.
(1981), the uncertainty in the profile width is estimated by
allowing the 20 and 50 percent flux density levels to vary within the
r.m.s.  noise of the profile.  The systemic velocity is, in principle,
determined as the error-weighted mean of the midpoints of the two
velocities at the 20 and 50 percent flux density levels.  Due to the low
signal-to-noise ratio of several global profiles, the 20 percent level
width, however, is often unreliable.  Therefore, this flux density level
has only been used in cases where the peak signal-to-noise ratio in the
profile is larger than 15; otherwise the systemic velocity relies only
on the 50 percent flux density level.



\section{Sample properties}

The sample of galaxies presented here is not complete in any
magnitude or diameter-limited sense.
Morphological types range from S0 to Im, with an over-abundance
of  Sb galaxies, and the absolute blue magnitudes range from -22 to -14
($H_0$ = 75 km
s$^{-1}$ Mpc$^{-1}$).
Table 1  presents the main results; the meaning of the columns is as
follows:

\noindent Column 1) UGC number.

\noindent Column 2) NGC or IC number.

\noindent Column 3) Spectral bandwidth of the observation. (Note
that 2.5$n$ MHz bandwidth at the 21-cm line rest frequency
 corresponds to 527$n$ km s$^{-1}$.)

\noindent Column 4) Hanning tapered velocity resolution.

\noindent Column 5) Morphological type as given in
the RC3.

\noindent Column 6) Inclination, determined as $cos\, i$ = $(q^2 -
q^2_0)/(1-q^2_0)$ where $i$ is the inclination; $q$ is the ratio
of the observed minor and major axis sizes at the 25th blue magnitude
isophote as given in the RC3.  The intrinsic axial ratio $q_0$ of the
disk has been  been taken from Guthrie
(1992) (0.4 for type S0; 0.2 for types S0/a, Sa and Sab; 0.1
for type Sb; 0.15 for type Sbc; 0.1 for types Sc, Scd and Sd; 0.4 for
types Sm and Im).   For  the  types S?, a value of $q_0$ = 0.2
has been taken.

\noindent Column 7) Distance in Mpc, taking $H_0$ equal to 75 km s$^{-1}$
Mpc$^{-1}$ and correcting the heliocentric radial velocity for the
motion of the Sun with respect to the centroid of the
Local Group, using ($V_{\rm hel}$ + 300
sin\,$l$$\cdot$cos\,$b$)/$H_0$
(de Vaucouleurs et al. 1976);
$V_{\rm hel}$ is the H\,{\sc i} systemic velocity (cf. Column 13)
or, if unavailable, the
optical systemic velocity.

\noindent Column 8)  Angular optical diameter at the 25th blue magnitude
isophote as given in the RC3.

\noindent Column 9) Apparent blue magnitude as given in the RC3,
corrected to a `face-on' appearance, and corrected for Galactic and
external extinction.  If not given in the RC3, this quantity has been
taken from the UGC and corrected according to the transformations
in the RC3.

\noindent Column 10) Observed width of the global profile at the 50
percent level of the flux density peak. No widths are given for
galaxies with poor global profiles. For widths at the 20 percent level,
see note {\sl c} to Table 1.

\begin{table*}
\begin{scriptsize}
\caption[]{Results of  short WSRT observations for 57 UGC galaxies.}
\begin{flushleft}
\begin {tabular}{r@{\ \ } l@{\ } c@{\ \ } c@{\ \ } l@{\ \ } c@{\ \ }
 c@{\ \ } c@{\ \ } c@{\ \ } r@{\ }@{$\pm$} @{\ } c@{\ } c@{\ }
 c@{\ } r@{\ } @{$\pm$} @{\ }  l@{\ } c@{\ } c@{\ \ } c@{\ } c@{\ }}
\hline
\noalign{\medskip}
UGC & NGC/IC&$B$&$\Delta V$&$Type$&$i$&$d$&$D_{25}$&$B^0_{\rm T}$&
\multicolumn{2}{c}{$W_{50}$}&
$W^i_{50}$&$V_{\rm opt}$&
\multicolumn{2}{c}{$V_{\rm HI}$}&$FI$&$M_{\rm HI}$&
$M_{\rm HI}/L_{\rm B}$&Notes\\
\noalign{\smallskip}
 & &(MHz) &(km/s)& & ($^{\rm o}$)&(Mpc) &($^{\prime}$) &(mag) &
\multicolumn{2}{c}{(km/s)}
&(km/s)
&(km/s)&
\multicolumn{2}{c}{(km/s)}&\multicolumn{1}{c}{(Jy km/s)}&(10$^9$ M$_{\odot}$)
&(M$_{\odot}$/L$_{\rm B,\odot}$)&  \\
\noalign{\smallskip}
(1)&$\,\,\,\,\,\,\,$(2)&(3)&(4)&$\,\,\,$(5) &(6) &(7) &(8) &(9) &
\multicolumn{2}{c}{(10)}&(11) &(12) &\multicolumn{2}{c}{(13)}
&(14) &(15) &(16)&(17)  \\
\noalign{\smallskip}
\hline
\noalign{\medskip}

  1034& N  \phantom{2}551&5.0&       34.1 &Sbc  & 66   &65.6&1.82& 12.69&
\multicolumn{2}{c}{\ }    &  &4901
&\multicolumn{2}{c}{\ }     &  \phantom{1}4.9   & 4.97 &0.09&a\\

  1291& I\,\, 1731& 2.5&    \, 8.4  &Sc   & 51   &44.8&1.55&
13.45&\multicolumn{2}{c}{\ }   &
&3426&\multicolumn{2}{c}{\ }
 &  \phantom{1}7.6   & 3.60 &0.29&\\

  1970&       &  5.0  &   16.7 &Scd  & 86   &23.7&2.34& 13.34& 209&
\phantom{1}7   &193 &1915 &1914&\phantom{1}4  & \, 5.3   & 0.70 &0.19&\\

  2067&       &  5.0  &   16.8 &Sab  & 90   &51.3&1.95& 13.57& 98 &
\phantom{1}8   &\phantom{1}87 &3839 &3885& \phantom{1}4  & 15.3   & 9.49
&0.66&b\\

  2069&       &  2.5  &   \, 8.4  &Sd   & 54   &49.9&2.34& 12.54& 257&   25
&298 &3715 &3777&  12 & 17.7   & 10.4 &0.30&\\

  2392&       &  5.0  &   16.7 &Scd  & 77   &21.4&1.86& 13.50& 148&   11  &137
&1548 &1551&  \phantom{1}6  &  \phantom{1}6.0   & 0.65 &0.24&\\

  2526&        &5.0  &   34.1 &Sb   & 80   &66.1&3.55& 11.67& 589&   32
&580&4996 &4960& 16&\phantom{1}6.9 &7.13&0.05& \\

  2548&  N 1207& 2.5 &  \, 8.4  &Sb   & 44   &64.1&2.29&
12.54&\multicolumn{2}{c}{\ }   & &4787 &\multicolumn{2}{c}{\ } &
\phantom{1}5.8   & 5.63 &0.10&\\

  2617&       &  2.5  &  \, 8.4  &Sd   & 70   &63.7&2.51&
12.48&\multicolumn{2}{c}{\ }  & &4860 &\multicolumn{2}{c}{\ } &  \phantom{1}5.1
  & 4.88 &0.08&\\

  2920&       &  5.0  &   34.0 &Scd  & 83   &53.1&2.29& 12.37& 385&   22  &369
&4158 &4151&  11 &  13.0    & 8.66 &0.19&\\

  3165&       &  5.0  &   16.8 &Im   & 61   &51.6&2.29& $\!\!\!$14.6& 101&
\phantom{1}3   &103 &3780 &3759&  \phantom{1}1  & 16.2   & 10.2 &1.73&c\\

  3458&       &  2.5  &  \, 8.4  &Sb   & 77   &59.4&2.40&
13.51&\multicolumn{2}{c}{\ }   & &4292 &\multicolumn{2}{c}{\ }   &
\phantom{1}3.4   & 2.83 &0.14&\\

  4271&  N 2523& 5.0 &   33.9 &Sbc  & 53   &47.6&2.95& 12.07& 448&   26  &540
&3415 &3468&   13&  12.8  &  6.84& 0.14&d\\

  4630&   I\,\,\phantom{2} 520& 5.0 &   33.9 &Sab  & 38   &42.8&1.95&
12.32&\multicolumn{2}{c}{\ }       & &3528 &\multicolumn{2}{c}{\ }   &
\phantom{1}4.0  &  1.73& 0.05&\\

  4705&  N 2710& 5.0 &   33.8 &Sb   & 61   &36.0&2.00& 13.14& 289&   13  &309
&2538 &2508 &  \phantom{1}7 &  15.4  &  4.71& 0.45&\\

  4825&  N 2748& 5.0 &   16.7 &Sbc  & 69   &23.4&3.02& 11.69& 288&
\phantom{1}6   &291 &1456 &1476 &  \phantom{1}2 &  29.4  &  3.79& 0.23&c\\

  4906&       &  5.0  &   33.8 &Sa   & 79   &32.2&2.00&
12.89&\multicolumn{2}{c}{\ }    & &2322 &\multicolumn{2}{c}{\ }   &
\phantom{1}4.9  &  1.20& 0.11&\\

  6348&  N 6348& 5.0 &   16.7 &Sb   & 72   &25.1&2.00& 13.19& 235&   10  &230
&1966 &1924&   \phantom{1}5 &  15.4  &  2.30& 0.47&\\

  6458&  N 3683& 5.0 &   16.7 &Sc   & 69   &20.4&1.86& 12.67& 339&   32  &346
&1656 &1722&   16&  20.8  &  2.04& 0.39&\\

  7127&  N 4133& 5.0 &   16.7 &Sb   & 42   &16.5&1.82& 12.78& 289&   37  &404
&1296 &1359&18 &   \phantom{1}8.0  &  0.52& 0.17&\\

  7290&  N 4220& 5.0 &   16.7 &S0   & 90   &14.2&3.89&
12.23&\multicolumn{2}{c}{\ }      &   &\, 979 &\multicolumn{2}{c}{\ }   &
\phantom{1}6.2  &  0.29& 0.08&\\

  7329&  N 4250& 5.0 &   33.7 &S0   & 42   &24.9&2.69& 12.70& 152&   16  &202
&2032 &2019&  \phantom{1}6 &  19.4  &  2.84& 0.38&c\\

  7443&  N 4314& 5.0 &   16.7 &Sa   & 28   &12.5&4.17&
11.21&\multicolumn{2}{c}{\ }   & &\, 963 &\multicolumn{2}{c}{\ }     &
\phantom{1}1.1  &  0.04& 0.01&\\

  7714&  N 4525& 5.0 &   16.7 &Scd  & 60   &14.9&2.57& 12.41& 149&
\phantom{1}9   &156 &1131 &1172 &  \phantom{1}4 &   \phantom{1}6.5  &  0.34&
0.10&\\

  7962&  N 4693& 5.0 &   16.7 &Sd   & 80   &23.7&2.46& 12.98& 256&
\phantom{1}9   &244 &1647 &1674 &  \phantom{1}5 &  14.1  &  1.86& 0.35&\\

  7994&  N 4750& 5.0 &   16.7 &Sab  & 25   &19.5&2.04& 12.07& 292&   22  &660
&1614 &1618 &  11&   \phantom{1}8.7  &  0.78& 0.09&\\

  8835&  N 5362& 5.0 &   33.8 &Sb   & 64   &29.0&2.29& 12.70& 296&
\phantom{1}8    &309 &2232 &2175 &  \phantom{1}4 &   \phantom{1}9.4  &  1.86&
0.18&\\

  8935&  N 5422& 5.0 &   16.8 &S0   & 90   &24.3&3.89&
12.79&\multicolumn{2}{c}{\ }    & &1782 &\multicolumn{2}{c}{\ }   &
\phantom{1}4.8  &  0.67& 0.10&\\

  8958&  N 5443& 5.0 &   16.7 &Sb   & 69   &24.1&2.69& 12.52& 408&   59  &420
&1921 &1787 &  29&   \phantom{1}5.9  &  0.81& 0.10&\\

  9115&  N 5526& 5.0 &   33.7 &Sbc  & 90   &29.1&1.78& $\!\!\!$14.5& 200&59
  &182 &1971 &2086&29     &   \phantom{1}8.1  &  1.62& 0.85&\\

  9428&  N 5707& 5.0 &   33.8 &Sab  & 88   &28.9&2.57& 12.49& 366&   24  &348
&2208 &2191 & 12&  11.6  &  2.29& 0.19&\\

  9516&  I\,\, 1056& 2.5 &  \, 8.4  &Sb   & 45   &53.3&1.82&
13.76&\multicolumn{2}{c}{\ }  & &4013 &\multicolumn{2}{c}{\ } &  10.3  &  6.91&
0.53&\\

  9852&  N 5929& 5.0 &   33.8 &Sb   & 59   &32.7&1.66&
$\!\!\!$13.0&\multicolumn{2}{c}{\ }   & &2672 & \multicolumn{2}{c}{\ } &
\phantom{1}3.8  &  0.96& 0.10&e\\

  9948&  N 5981& 5.0 &   33.7 &Sc   & 83   &28.1&2.82&
12.79&\multicolumn{2}{c}{\ }  &  &1764 &\multicolumn{2}{c}{\ } &
\phantom{1}0.9  &  0.17& 0.02&\\

 10437&       &  5.0  &   33.7 &S?   & 30   &32.4&2.09& 14.57& 211&   11  &385
&2595 &2602&  \phantom{1}6 &  14.8  &  3.68& 1.61&\\

 10713&       &  5.0  &   33.6 &Sb   & 81   &15.2&1.82& 12.79& 228&   38  &212
&1112 &1074& 19&  19.6  &  1.06& 0.41&\\

 10917&  I\,\, 1265& 5.0 &   16.7 &Sab  & 67   &24.9&2.04& 13.45& 271&   36
&276 &2161 &2156 &18&   \phantom{1}6.8  &  0.99& 0.26&\\

 11432&  N 6796& 5.0 &   16.7 &Sbc  & 79   &28.1&1.86& 12.26& 442&   34  &433
&2090 &2198  & 17&  13.2  &  2.47& 0.17&\\

 11466&       &  5.0  &   33.6 &S?   & 56   &11.2&2.00& 11.33& 167&
\phantom{1}7   &181 &\, 782 &818  & \phantom{1}2  &  33.3  &  0.98&0.18&c\\

 11470&  N 6824& 5.0 &   16.8 &Sb   & 47   &42.4&1.70&
11.89&\multicolumn{2}{c}{\ }    & &3337 &\multicolumn{2}{c}{\ }   &
\phantom{1}4.6  &  1.95& 0.04&\\

 11635&       &  2.5  &  \, 8.4  &Sbc  & 66   &60.7&2.88&
12.88&\multicolumn{2}{c}{\ }     &  &4731 &\multicolumn{2}{c}{\ }       &
\phantom{1}6.0  &  5.22& 0.14&\\

 11920&       &  5.0  &   33.6 &S0/a & 52   &15.3&2.40&
11.12&\multicolumn{2}{c}{\ }     & & 1145&\multicolumn{2}{c}{\ }  &
\phantom{1}5.3&   0.29& 0.02&\\
 & & & & & & & & &\multicolumn{2}{c}{\ } & & &\multicolumn{2}{c}{\ } & & & & \\

  2521&  N 1186& 2.5 &  \, 8.4  &Sbc  & 69   &37.0&3.16&
10.89&\multicolumn{2}{c}{\ }         &   &2658&\multicolumn{2}{c}{\ } & &
$\!\!\!\!\!<$1.13&$\!\!\!\!<$0.01\\

  3994&   I\,\,\phantom{2} 469& 5.0 &   33.7 &Sab  & 65
&25.5&2.19&12.87&\multicolumn{2}{c}{\ }    &
&
2080&\multicolumn{2}{c}{\ } & &$\!\!\!\!\!<$0.63&$\!\!\!\!<$0.09\\

  4544&  N 2639& 5.0 &   33.9 &Sa   & 54   &42.1&1.82&
12.26&\multicolumn{2}{c}{\ }  &          &3198&\multicolumn{2}{c}{\ } &
&$\!\!\!\!\!<$0.76&$\!\!\!\!<$0.02\\

  4645&  N 2681& 5.0 &   33.6 &S0/a & 25   &10.4&3.63&
10.90&\multicolumn{2}{c}{\ }   &       &\, 683&\multicolumn{2}{c}{\ } &
&$\!\!\!\!\!<$0.03&$\!\!\!\!<$0.01\\

  5568&  N 3182& 5.0 &   33.7 &Sa   & 32   &30.2&1.82&
12.88&\multicolumn{2}{c}{\ }  &         &2130&\multicolumn{2}{c}{\ } &
&$\!\!\!\!\!<$0.53&$\!\!\!\!<$0.06\\

  5692&       &  5.0  &   16.6 &Sm   & 66   &\, 2.7 &3.24&
13.24&\multicolumn{2}{c}{\ }  &         &\, 180&\multicolumn{2}{c}{\ } &
&$\!\!\!\!\!<$0.01&$\!\!\!\!<$0.03\\

  6392&  N 3652& 5.0 &   33.7 &Scd  & 72   &31.1&2.00&
12.15&\multicolumn{2}{c}{\ }  &         &2096&\multicolumn{2}{c}{\ } &
&$\!\!\!\!\!<$0.23&$\!\!\!\!<$0.01\\

  6484& N 3683A& 5.0 &   33.8 &Sc   & 44   &28.5&2.34&
12.38&\multicolumn{2}{c}{\ }  &         &2446&\multicolumn{2}{c}{\ } &
&$\!\!\!\!\!<$0.27&$\!\!\!\!<$0.02\\

  7328&  N 4245& 5.0 &   16.7 &S0/a & 42   &\, 9.2 &2.88&
12.05&\multicolumn{2}{c}{\ }  &        &\, 890&\multicolumn{2}{c}{\ } &
&$\!\!\!\!\!<$0.04&$\!\!\!\!<$0.02\\

  7429&  N 4319& 5.0 &   16.7 &Sab  & 40   &23.3&2.95&
12.55&\multicolumn{2}{c}{\ } &         &1700&\multicolumn{2}{c}{\ } &
&$\!\!\!\!\!<$0.32&$\!\!\!\!<$0.04\\

  7572&  N 4441& 5.0 &   16.7 &S0   & 44   &34.4&3.24&
13.25&\multicolumn{2}{c}{\ }  &         &1466&\multicolumn{2}{c}{\ } &
&$\!\!\!\!\!<$0.63&$\!\!\!\!<$0.07\\

  8722&  N 5308& 5.0 &   33.7 &S0   & 90   &27.5&3.72&
12.43&\multicolumn{2}{c}{\ }  &         &2041&\multicolumn{2}{c}{\ } &
&$\!\!\!\!\!<$0.42&$\!\!\!\!<$0.04\\

  9016&  N 5475& 5.0 &   16.7 &Sa   & 80   &22.2&2.04&
12.88&\multicolumn{2}{c}{\ } &         &1720&\multicolumn{2}{c}{\ } &
&$\!\!\!\!\!<$0.41&$\!\!\!\!<$0.08\\

 10560&  I\,\, 1231& 2.5 &  \, 8.5  &Scd  & 63   &67.8&2.19&
13.07&\multicolumn{2}{c}{\ }  &         &5107&\multicolumn{2}{c}{\ } &
&$\!\!\!\!\!<$2.70&$\!\!\!\!<$0.07\\

 10676&       &  5.0  &   33.6 &Im   & 70   &\, 9.7 &1.82&
$\!\!\!$15.9&\multicolumn{2}{c}{\ }  &    &\, 524&\multicolumn{2}{c}{\ } &
&$\!\!\!\!\!<$0.01&$\!\!\!\!<$0.17\\
\hline
\noalign{\medskip}
\multicolumn{19}{l}{\footnotesize a. (UGC 1034) Possibly also HI
emission at the edge of the spectral band.}\\
\multicolumn{19}{l}{\footnotesize b. (UGC 2067) The observed global
profile is possibly that of the companion UGC 2065, a face-on Sm}\\
\multicolumn{19}{l}{\footnotesize \phantom{b.}
galaxy at 2.5 arcmin  from UGC 2067. }\\
\multicolumn{19}{l}{\footnotesize c. W$_{20}$ has
been determined for four galaxies: UGC 3165:
141 ($\pm$5) km
s$^{-1}$;
UGC 4825: 319 ($\pm$8) km s$^{-1}$;}\\
\multicolumn{19}{l}{\footnotesize \phantom{c.} UGC 7329: 236 ($\pm$17) km
s$^{-1}$; UGC 11466:
251 ($\pm$7) km
s$^{-1}$.}\\
\multicolumn{19}{l}{\footnotesize d. (UGC 4271) Arp 9.}\\
\multicolumn{19}{l}{\footnotesize e. (UGC 9852) Arp 90.}
\end{tabular}
\end{flushleft}
\end{scriptsize}
\end{table*}

\noindent Column 11) Width of the global profile at a level of 50
percent of the flux density peak, corrected for inclination and for
instrumental and turbulent broadening.  The instrumental broadening has
been determined  using the precepts of Bottinelli et al.
(1990).  The correction for the turbulent motions has been
 derived using the prescriptions of Tully \& Fouqu\'{e} (1985).
The value for the contribution by non-circular motion is taken as 14 km
s$^{-1}$ (Bottinelli et al. 1983).

\noindent Column 12) Optical heliocentric  systemic velocity
from the RC3.

\noindent Column 13) 21-cm heliocentric  systemic velocity
derived from the global H\,{\sc i} profile.

\noindent Column 14) Integral of the global profile.  The
uncertainties are of the order 10 - 20 percent, but  may
be larger for  profiles
with low signal-to-noise ratio.

\noindent Column 15) Total H\,{\sc i} mass, calculated as $M_{\, \rm
HI}$
= $2.36  \cdot 10^5 \cdot d^2 \cdot FI\,\, M_{\odot}$ (assuming that the
H\,{\sc i} is
optically thin); $d$ is the distance in Mpc and $FI$ is the flux
integral in Jy km s$^{-1}$.  For the non-detections, listed at the
bottom of Table 1, we estimate a
2$\sigma$ upper limit for the H\,{\sc i} mass, assuming typical values for the
corrected profile widths for different morphological types (200 km
s$^{-1}$ for types S0 to
Sc;  120 km s$^{-1}$ for type Scd;  70 km s$^{-1}$
for type Sm;  50 km s$^{-1}$ for type Im).

\noindent Column 16) Total H\,{\sc i} mass-to-blue luminosity ratio.

\noindent Column 17) Notes on individual galaxies.

\section{Discussion}

For the galaxies presented here, H\,{\sc i} information in the RC3 is
absent.  As some of them are fairly large ($d$$_{\rm b}$ $>$ 3 arcmin),
we suspect that several galaxies may have been observed earlier but have
not been detected.  It is therefore of some interest to look at the
H\,{\sc i} content of the galaxies in our sample.

\begin{figure}
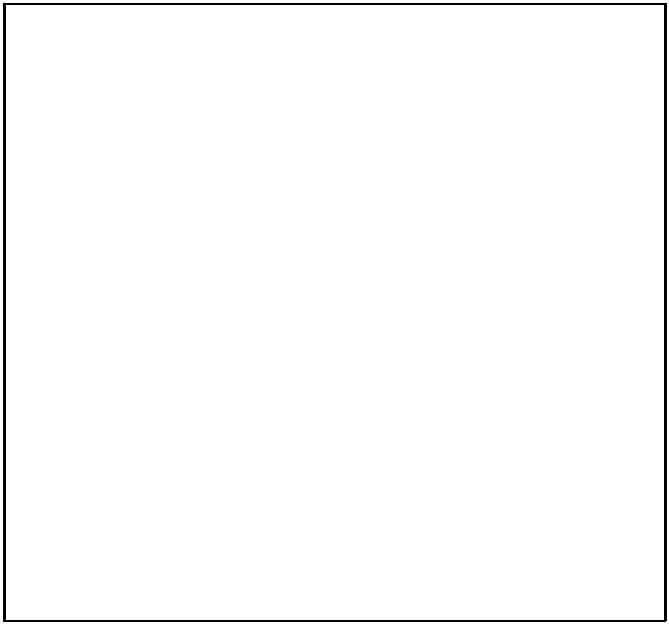

\picplace{8.2 cm}
\caption[]{H\,{\sc i} mass-to-blue luminosity ratio versus
morphological type. The solid line represents the mean value for the
RC3-UGC
sample from Roberts \& Haynes (1994);
the dashed lines are the 25 and 75 percentiles. The arrows indicate
upper limits for the non-detections.}
\end{figure}

In Figure 2 we plot the H\,{\sc i} mass-to-blue luminosity ratio versus
morphological type.  Arrows indicate upper limits.  The solid line
represents the mean relation, and the dashed lines 25 and 75
percentiles, as given by Roberts \& Haynes (1994).  The graph shows that
 the H\,{\sc i} content is somewhat below the mean relation, especially
for later types.
A  number of our
upper limits are interestingly low: a factor of 20 below the mean.  This
indicates that we are now exploring the fainter part of the
H\,{\sc i} mass function.  Apparently, for most galaxies in the RC3 with
a high H\,{\sc i} content an H\,{\sc i} entry is available.

The Tully-Fisher relation (absolute blue magnitude versus corrected profile
width) is shown in Figure 3.  Except for UGC 2067 (whose measurement
probably refers to the Sm galaxy UGC 2065), there is fair agreement with
the mean relation given by Pierce \& Tully (1992).  Note that the three
galaxies furthest below the line are close to face-on and their
inclination corrected profile widths are therefore uncertain.

\begin{figure}
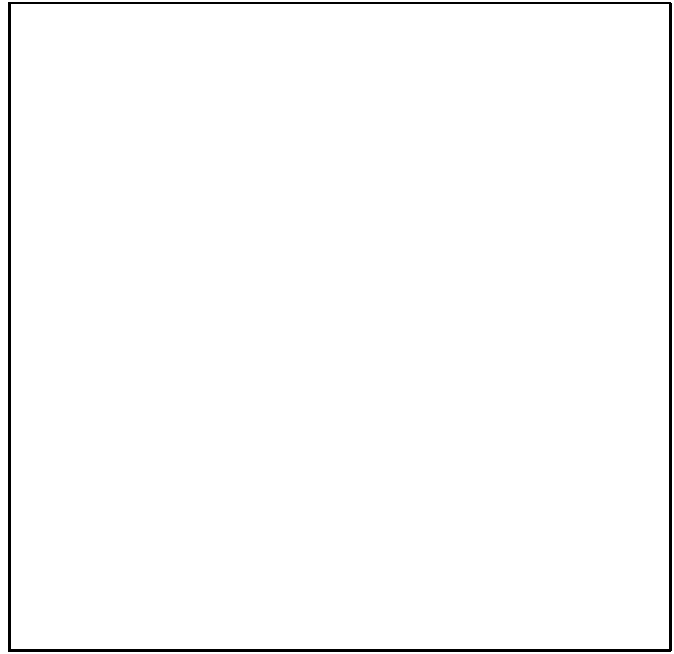

\picplace{8.6 cm}
\caption[]{Absolute blue magnitude versus profile width.
The solid line is the Tully-Fisher relation as
given by Pierce \& Tully  (1992, corrected to $H_0$ = 75 km s$^{-1}$
Mpc$^{-1}$). The measurement for UGC 2067 probably refers to the Sm
galaxy UGC 2065 at 2.5 arcmin from UGC 2067.}
\end{figure}

\begin{acknowledgements}

The WSRT is operated by the Netherlands Foundation for Research in
Astronomy with financial support from the Netherlands Organisation for
Scientific Research (NWO). We thank J.M. van der Hulst and R. Sancisi
for comments.

\end{acknowledgements}

{}

\end{document}